\documentclass[english,prd,superscriptaddress,nofootinbib,preprintnumbers,eqsecnum]{revtex4}
\usepackage{graphicx}
\usepackage{dcolumn}
\usepackage{bm}
\usepackage{latexsym}
\usepackage{amsfonts}
\usepackage{amssymb}
\usepackage{amsmath}
\usepackage{mathrsfs}
\usepackage{color}

\def\be{\begin{equation}}
\def\ee{\end{equation}}
\def\ba{\begin{eqnarray}}
\def\ea{\end{eqnarray}}

\begin{document}

\title{Cosmology in generalized Horndeski theories 
with second-order equations of motion}

\author{Ryotaro Kase}
\affiliation{Department of Physics, Faculty of Science, Tokyo University of Science, 1-3,
Kagurazaka, Shinjuku, Tokyo 162-8601, Japan}

\author{Shinji Tsujikawa}
\affiliation{Department of Physics, Faculty of Science, Tokyo University of Science, 1-3,
Kagurazaka, Shinjuku, Tokyo 162-8601, Japan}

\date{\today}

\begin{abstract}

We study the cosmology of an extended version of Horndeski theories 
with second-order equations of motion on the flat 
Friedmann-Lema\^{i}tre-Robertson-Walker (FLRW) background. 
In addition to a dark energy field $\chi$ associated with the gravitational sector, 
we take into account multiple 
scalar fields $\phi_I$ ($I=1,2\cdots,N-1$) characterized by the 
Lagrangians $P^{(I)}(X_I)$ with $X_I=\partial_{\mu}\phi_I\partial^{\mu}\phi_I$. 
These additional scalar fields can model the perfect fluids of radiation 
and non-relativistic matter. 
We derive propagation speeds of scalar and tensor perturbations 
as well as conditions for the absence of ghosts. 
The theories beyond Horndeski induce non-trivial modifications
to all the propagation speeds of $N$ scalar fields,  
but the modifications to those for the matter fields $\phi_I$ 
are generally suppressed relative to that for the dark energy field $\chi$.
We apply our results to the covariantized Galileon with 
an Einstein-Hilbert term in which partial derivatives 
of the Minkowski Galileon are replaced by covariant derivatives. 
Unlike the covariant Galileon with second-order equations 
of motion in general space-time, the scalar 
propagation speed square $c_{s1}^2$ associated with the 
field $\chi$ becomes negative during the matter era
for late-time tracking solutions, so the two Galileon 
theories can be clearly distinguished at the level of 
linear cosmological perturbations.

\end{abstract}

\maketitle




\section{Introduction}
\label{intro} 

One of the most serious mysteries in modern cosmology is dark energy--
the shadowy source for the late-time cosmic acceleration \cite{Riess,Perlmutter}. 
The cosmological constant remains the most economical explanation 
for the origin of dark energy \cite{Lambda}, but there are other alternatives such as 
scalar fields and modifications of Einstein gravity \cite{review}. 
Interestingly, the recent combined analysis based on the observations of 
Supernovae type Ia (Sn Ia), Cosmic Microwave Background (CMB), and 
Baryon Acoustic Oscillations (BAO) showed that the cosmological 
constant is in mild tension with the data \cite{WMAP,Planck}.

Many dark energy models proposed in the literature involve at least one 
scalar degree of freedom. Typical examples are quintessence \cite{quin} 
and k-essence \cite{kes}, in which the potential energy and the kinetic energy 
of a minimally coupled scalar field $\chi$ drives the cosmic acceleration. 
Even in the case of $f(R)$ gravity where the Lagrangian 
contains non-linear terms of the Ricci scalar 
$R$ \cite{fR1,fRviable}, there exists a gravitational 
scalar degree of freedom coupled to non-relativistic 
matter \cite{APT}.

If the derivatives higher than second order appear in the equations 
of motion, the corresponding theory is usually prone to a 
ghost-like (Ostrogradski) instability \cite{Ostro}
related with the Hamiltonian unbounded from below. 
The Lagrangian of most general scalar-tensor theories with 
second-order equations of motion free from the Ostrogradski 
instability was first derived by Horndeski in 1973 \cite{Horndeski}.
The Horndeski theories accommodate a wide range of 
Lorentz-invariant dark energy models with one scalar 
degree of freedom--including quintessence \cite{quin}, 
k-essence \cite{kes}, $f(R)$ gravity \cite{fR1,fRviable}, 
Brans-Dicke theory \cite{Brans}, 
and Galileons \cite{Nicolis,Galileon}.

On the flat isotropic Friedmann-Lema\^{i}tre-Robertson-Walker (FLRW) 
background, it is also possible to provide a unified framework of 
modified gravitational theories based on the effective 
field theory (EFT) of cosmological perturbations \cite{Weinberg}-\cite{Xian}.
The building block of this approach is a general action in unitary gauge  
that depends on the lapse function $N$ and geometric scalar 
quantities (those from extrinsic and intrinsic curvatures)
constructed in the 3+1 Arnowitt-Deser-Misner (ADM) 
formalism \cite{ADM}. 
Expansion of the action up to quadratic order in perturbations shows 
that there exist spatial derivatives higher than 
second order in linear perturbation equations of motion, 
while time derivatives remain of second order. 
Gleyzes {\it et al.} \cite{Fedo} derived conditions for the absence of 
higher-order spatial derivatives, which are in fact satisfied 
for Horndeski theories.

The Horndeski Lagrangian contains four arbitrary functions 
$G_i (\chi,Y)$ ($i=2,3,4,5$), where $Y=\partial_{\mu}\chi 
\partial^{\mu}\chi$ is the kinetic energy of a scalar 
field $\chi$ \cite{Deffayet,Toni,KYY}. 
In unitary gauge this Lagrangian can be expressed in terms of 
three-dimensional ADM scalar quantities 
with four functions $A_i(t,N)$ ($i=2,3,4,5$) 
and two functions $B_i(t,N)$ ($i=4,5$), where 
the dependence of $\chi$ and $Y$ in the functions 
$G_i$ translate to that of time $t$ and lapse $N$ 
in $A_i$ and $B_i$ \cite{Fedo}.
In Horndeski theories the functions $B_{4}$ and 
$B_{5}$ are related with $A_{4}$ and $A_5$ according to 
$A_4=2YB_{4Y}-B_4$ and $A_5=-YB_{5Y}/3$, 
where $B_{iY} \equiv \partial B_i/\partial Y$ \cite{Gleyzes}.

Even for the generalized version of Horndeski theories 
in which $B_4$ and $B_5$ are not necessarily related 
with $A_4$ and $A_5$, 
Gleyzes, Langlois, Piazza, and Vernizzi (GLPV) showed 
that, on the flat FLRW background, the perturbation 
equations of motion are of second order 
with one scalar propagating 
degree of freedom \cite{Gleyzes}. 
This second-order property also holds for the odd-type 
perturbations on the spherically symmetric and 
static background \cite{KGT}. 
In GLPV theories, the presence of symmetries 
in space-time allows for the absence of derivatives 
higher than quadratic order.

The covariantized version of the original Galileon \cite{Nicolis}--whose Lagrangian 
is derived by replacing partial derivatives of the Minkowski Galileon \cite{Nicolis} 
with covariant derivatives-- belongs to a class of GLPV theories \cite{Gleyzes}. 
This is different from the covariant Galileon \cite{Galileon} in which gravitational 
counter terms are added to eliminate derivatives higher than 
second order in general space-time. In other words, the covariant 
Galileon falls in a class of Horndeski theories, while the covariantized 
Galileon does not. 

In order to study the cosmological dynamics based on GLPV theories, we need to 
take into account matter fields (such as non-relativistic matter and 
radiation) other than the scalar field $\chi$ responsible for dark energy. 
In the presence of an additional scalar field $\phi$ with a kinetic energy 
$X=\partial_{\mu}\phi \partial^{\mu}\phi$, the conditions for eliminating 
derivatives higher than second order have been derived in Ref.~\cite{LS} 
for the action depending on $\phi$ and $X$ 
as well as on other ADM scalar quantities.
In Ref.~\cite{LS} the authors also obtained conditions for the avoidance 
of ghosts and Laplacian instabilities associated with scalar and tensor 
perturbations. In GLPV theories it was recognized that the matter
propagation speed $c_m$ is affected by the scalar degree of freedom 
$\chi$ \cite{Gleyzes}, but this is not the case for Horndeski 
theories \cite{deFe,LS}.

In this paper, we derive propagation speeds of scalar and tensor 
perturbations as well as no-ghost conditions in GLPV theories
in the presence of multiple matter fields on the flat FLRW background.
In addition to the dark energy field $\chi$, we take into account 
scalar fields $\phi_I$ ($I=1,2,\cdots,N-1$) with the 
Lagrangians $P^{(I)}(X_I)$ depending on 
$X_I=\partial_{\mu}\phi_I\partial^{\mu}\phi_I$. 
This prescription can accommodate the perfect fluids of radiation 
and non-relativistic matter \cite{Hu,Mukoh}.
We obtain an algebraic equation for the propagation speeds 
of multiple scalar fields and estimate to what extent the 
difference arises by going beyond Horndeski theories.

We then apply our results to two different theories-- covariantized Galileon 
and covariant Galileon. Since the difference between them
appears only in the functions $B_i$ but not in the functions $A_i$, 
the background equations of motion for the covariantized Galileon 
are exactly the same as those for the covariant Galileon. 
At the level of perturbations, however, these two theories can be 
clearly distinguished from each other.
For the covariantized Galileon the propagation speed square 
$c_{s1}^2$ of the field $\chi$ becomes 
negative in the deep matter era for late-time tracking solutions, 
whereas in the covariant Galileon it remains positive. 
We also show that the matter sound speed squares
of the fields $\phi_I$ for the covariantized Galileon 
are similar to those for the covariant Galileon.

Our paper is organized as follows.
In Sec.~\ref{theorysec} we briefly review GLPV theories and derive 
the background equations of motion on the flat FLRW space-time.
In Sec.~\ref{persec} we obtain the second-order action for scalar/tensor 
perturbations in GLPV theories with multiple scalar fields. 
We derive not only no-ghost conditions but also an $N$-th order 
algebraic equation for the scalar propagation speed squares $c_s^2$.
In Sec.~\ref{Galisec} we study the cosmology based on the two 
Galileon theories (covariantized and covariant Galileons). 
We discuss how these theories can be distinguished from each other, 
paying particular attention to the evolution of the 
scalar propagation speeds.
Sec.~\ref{consec} is devoted to conclusions.


\section{GLPV theories and the background equations 
of motion on the flat FLRW background}
\label{theorysec} 

We employ the $3+1$ decomposition in the ADM 
formalism described by the line element 
\be
ds^{2}=g_{\mu \nu} dx^{\mu} dx^{\nu}
=-N^{2}dt^{2}+h_{ij}(dx^{i}+N^{i}dt)(dx^{j}+N^{j}dt)\,, 
\label{ADMmetric}
\ee
where $N$ is the lapse, $N^{i}$ is the shift vector, and $h_{ij}$ 
is the three-dimensional metric. 
A unit normal vector orthogonal to constant time hypersurfaces 
$\Sigma_t$ is given by $n_{\mu}=(-N,0,0,0)$ with the 
normalization $n_{\mu} n^{\mu}=-1$.
The extrinsic curvature of $\Sigma_t$ is defined by 
\be
K_{\mu \nu} \equiv 
h_{\mu}^{\lambda} h_{\nu}^{\sigma}\,n_{\sigma ;\lambda}
=n_{\nu ;\mu}+n_{\mu} n^{\lambda}n_{\nu;\lambda}\,,
\label{Kmunu}
\ee
where a semicolon represents a covariant derivative.
In the second equality of Eq.~(\ref{Kmunu}) we have used 
the fact that the three-dimensional metric $h_{\mu \nu}$ can be 
expressed as $h_{\mu \nu}=g_{\mu \nu}+n_{\mu}n_{\nu}$. 
The internal geometry of the hypersurfaces is characterized 
by the three-dimensional Ricci tensor 
${\cal R}_{\mu \nu} \equiv {}^{(3)} R_{\mu \nu}$.

The EFT of cosmological perturbations advocated in 
Refs.~\cite{Weinberg}-\cite{Fedo} is based on the 
combination of geometric scalar quantities:
\be
K\equiv {K^{\mu}}_{\mu}\,,\qquad \mathcal{S}\equiv K_{\mu \nu }
K^{\mu \nu}\,,\qquad \mathcal{R}\equiv {\mathcal{R}^{\mu }}_{\mu }\,,
\qquad \mathcal{Z} \equiv \mathcal{R}_{\mu \nu}
\mathcal{R}^{\mu \nu }\,,\qquad \mathcal{U}
\equiv \mathcal{R}_{\mu \nu }K^{\mu \nu }\,,  
\label{threedef}
\ee
as well as the lapse $N$. 
We assume the existence of a scalar degree of freedom 
$\chi$ with the kinetic energy $Y \equiv 
g^{\mu \nu} \partial_{\mu} \chi \partial_{\nu} \chi$.
We choose the unitary gauge in which the 
constant field hypersurfaces coincide with the 
constant time hypersurfaces.
Since $\chi=\chi(t)$ and $Y=-N^{-2} \dot{\chi}(t)^2$, 
the $\chi$ and $Y$ dependence in the Lagrangian 
$L$ can be interpreted as the $N$ and $t$ dependence.

Expanding the Lagrangian $L(N,K,{\cal S},{\cal R},{\cal Z},{\cal U};t)$ 
up to second order of perturbations on the flat FLRW background 
with the line-element $ds^2=-dt^2+a^2(t) \delta_{ij}dx^idx^j$, 
Gleyzes {\it et al.} \cite{Fedo} showed that the {\it linear} 
perturbation equations of motion are of second order 
under the conditions
\ba
& &
L_{KK}+4HL_{{\cal S}K}+4H^2 L_{{\cal S}{\cal S}}+2L_{\cal S}=0\,,
\label{spacon1} \\
& &
L_{K{\cal R}}+2HL_{{\cal S}{\cal R}}+\frac12 L_{\cal U}
+HL_{K{\cal U}}+2H^2L_{{\cal S}{\cal U}}=0\,,
\label{spacon2} \\
& & 
4 \left( L_{{\cal R}{\cal R}}+2HL_{{\cal R} {\cal U}}+H^2L_{\cal UU} \right)
+3L_{\cal Z}=0\,,
\label{spacon3}
\ea
where $H \equiv \dot{a}/a$ is the Hubble parameter (a dot represents 
the derivative with respect to the time $t$), and a lower index 
of $L$ denotes the partial derivatives with respect to the scalar quantities, 
e.g., $L_{\cal S}=\partial L/\partial {\cal S}$.
Equations (\ref{spacon1})-(\ref{spacon3}) are sufficient conditions 
for eliminating spatial derivatives higher than second order.

Let us consider four-dimensional Horndeski theories described 
by the Lagrangian
\ba
L &=&
G_2(\chi,Y)+G_3(\chi,Y) \square \chi
+G_4 (\chi,Y)R-2G_{4Y} (\chi,Y) 
\left[ (\square \chi)^2-\chi^{;\mu \nu} \chi_{;\mu \nu} \right] \nonumber \\
&& +G_5(\chi,Y)G_{\mu \nu} \chi^{;\mu \nu}
+G_{5Y} (\chi,Y) \left[ (\square \chi)^3-3(\square \chi) 
\chi_{;\mu \nu} \chi^{;\mu \nu}+2\chi_{;\mu \nu}
\chi^{;\mu \sigma} {\chi^{;\nu}}_{;\sigma}\right]/3\,,
\label{LH}
\ea
where $G_i$ ($i=2,3,4,5$) are generic functions of $\chi$ and $Y$ 
with $G_{iY}= \partial G_{i}/\partial Y$,
$R$ and $G_{\mu \nu}$ are the Ricci scalar and the Einstein tensor 
in four dimensions. In terms of the ADM scalar quantities 
the Lagrangian (\ref{LH}) can be expressed as  \cite{Fedo}
\be
L=A_2+A_3K+A_4 (K^2-{\cal S})+B_4{\cal R}
+A_5 K_3+B_5 \left( {\cal U}-K {\cal R}/2 \right)\,,
\label{LH2}
\ee
where 
\ba
& & A_2=G_2-YF_{3\chi}\,,\qquad 
A_3=2(-Y)^{3/2}F_{3Y}-2\sqrt{-Y}G_{4\chi}\,,\qquad
A_4=2YG_{4Y}-G_4+YG_{5\chi}/2\,,\nonumber \\
& & B_4=G_4+Y(G_{5\chi}-F_{5\chi})/2\,,\qquad
A_5=-(-Y)^{3/2}G_{5Y}/3\,,\qquad
B_5=-\sqrt{-Y}F_{5}\,.
\label{AB}
\ea
The lower indices $\chi$ and $Y$ represent the partial derivatives 
with respect to $\chi$ and $Y$, respectively.
We have introduced two auxiliary functions $F_3$ and $F_5$ 
satisfying $G_3=F_3+2YF_{3Y}$ and 
$G_{5Y}=F_5/(2Y)+F_{5Y}$. 
The quantity $K_3$ in the Lagrangian (\ref{LH2}) is defined by 
\ba
K_3 &\equiv& K^3-3KK_{ij}K^{ij}+2K_{ij}K^{il}{K^{j}}_l \nonumber \\
&=& 3H(2H^2-2KH+K^2-{\cal S})+O(3)\,,
\ea
where the second equality is valid up to quadratic order in perturbations.
{}From Eq.~(\ref{AB}) the coefficients $A_4$, $B_4$, $A_5$, $B_5$ 
in Horndeski theories are related with each other as 
\be
A_4=2Y B_{4Y}-B_4\,,\qquad 
A_5=-YB_{5Y}/3\,.
\label{Hocon}
\ee

The GLPV theories \cite{Gleyzes} correspond to the Lagrangian (\ref{LH2}) 
in which the conditions (\ref{Hocon}) do not necessarily hold.
It is clear that the Lagrangian (\ref{LH2}) satisfies the three conditions 
(\ref{spacon1})-(\ref{spacon3}) even without the restriction (\ref{Hocon}), 
so the linear perturbation equations of motion on the FLRW background 
do not contain derivatives higher than second order.

In the presence of an additional matter fluid, we derive 
the background equations of motion for the action
\be
S=\int d^4 x \sqrt{-g}\,L+S^{M}\,, 
\label{Lag}
\ee
where $L$ is given by Eq.~(\ref{LH2}).
In unitary gauge, the functions $A_i$ and $B_i$ depend on 
$t$ and $N$ through the relations $\chi=\chi(t)$ and 
$Y=-N^{-2}\dot{\chi}(t)^2$. 
$S^{M}$ is the matter action with energy density 
$\rho_M$ and pressure $P_M$.

On the flat FLRW background, the equations of motion can be 
derived by expanding the action (\ref{Lag}) up to first order 
in perturbations and by varying the first-order action 
in terms of $\delta N$ and $\delta \sqrt{h}$, where 
$h$ is the determinant of the three-dimensional 
metric $h_{ij}$ \cite{Fedo}. 
They are given, respectively, by 
\ba
& & \bar{L}+L_N-3H{\cal F}=\rho_M\,,
\label{back1} \\
& & \bar{L}-\dot{\cal F}-3H {\cal F}=-P_M\,,
\label{back2}
\ea
where $\bar{L}$ is the background value of $L$, and 
${\cal F} \equiv L_K+2HL_{\cal S}$.
On using the properties $\bar{K}=3H$, $\bar{{\cal S}}=3H^2$, 
$\bar{K}_3=6H^3$, and $\bar{{\cal R}}=\bar{{\cal U}}=0$, 
Eqs.~(\ref{back1}) and (\ref{back2}) read
\ba
\hspace{-0.5cm}
& & A_2-6H^2 A_4-12H^3 A_5+2\dot{\chi}^2 
\left( A_{2Y}+3H A_{3Y}+6H^2 A_{4Y}+6H^3 A_{5Y} \right)=\rho_M\,,
\label{back1d} \\
\hspace{-0.5cm}
& & A_2-6H^2 A_4-12H^3 A_5-\dot{A}_3-4\dot{H}A_4
-4H\dot{A}_4-12H \dot{H}A_5-6H^2 \dot{A}_5=-P_M\,.
\label{back2d} 
\ea
Substituting the functions $A_i$ of Eq.~(\ref{AB}) into 
Eqs.~(\ref{back1d})-(\ref{back2d}), we reproduce the background 
equations of motion in Horndeski theories \cite{KYY,DKT} 
derived by the direct variation of the action (\ref{Lag}) with (\ref{LH}).

Equations (\ref{back1d}) and (\ref{back2d}) do not contain 
the functions $B_4$ and $B_5$.
This means that, at the background level, the theories with 
same values of $A_2, A_3, A_4, A_5$ but with different values 
of $B_4$ and $B_5$ cannot be distinguished from each other.  
In fact, this happens for the covariantized Galileon and the 
covariant Galileon mentioned in Introduction.
However, it is possible to distinguish between such 
theories at the level of perturbations. 
We shall address this issue in Sec.~\ref{Galisec}.


\section{Cosmological perturbations and propagation speeds of tensor 
and scalar modes}
\label{persec} 

In this section, we derive no-ghost conditions and scalar propagation speeds 
for the theory described by the Lagrangian $L(N,K,{\cal S},{\cal R},{\cal Z},{\cal U};t)$
in the presence of multiple scalar fields $\phi_I$ ($I=1,2,\cdots,N-1$). 
As we already mentioned, we choose the unitary gauge in which 
the perturbation of the field $\chi$ vanishes ($\delta \chi=0$).
The k-essence Lagrangian $P^{(I)} (X_I)$ with a kinetic energy 
$X_I \equiv g^{\mu \nu}\partial_{\mu}\phi_I \partial_{\nu}\phi_I$ 
can describe the perturbation of a barotropic 
perfect fluid \cite{Hu,Mukoh,LS}.
Let us then consider the theory with $N$ scalar fields 
($\chi$ and $\phi_1, \cdots, \phi_{N-1}$) given by the action 
\be
S=\int d^4 x \sqrt{-g} \left[ L(N,K,{\cal S},{\cal R},{\cal Z},{\cal U};t)
+\sum_{I=1}^{N-1}  P^{(I)}(X_I) \right]\,,
\label{Lagper}
\ee
which covers the theory (\ref{Lag}) with (\ref{LH2}) as a special case. 
The energy density $\rho^{(I)}$ and the equation of state $w_I$ 
of the scalar field $\phi_I$ are given, respectively, by 
\be
\rho^{(I)}=2X_I P^{(I)}_{X_I}-P^{(I)}\,,\qquad
w_I=\frac{P^{(I)}}{2X_I P^{(I)}_{X_I}-P^{(I)}}\,,
\label{rhowI}
\ee
where $P^{(I)}_{X_I}=\partial P^{(I)}/\partial X_I$.
Then, the total energy density $\rho_M$ and the pressure $P_M$ 
of the scalar fields $\phi_1, \cdots, \phi_{N-1}$ read
\be
\rho_M=\sum_{I=1}^{N-1} \left[2X_I P^{(I)}_{X_I}-P^{(I)} \right]\,,
\qquad
P_M=\sum_{I=1}^{N-1} P^{(I)}\,.
\ee
Combining Eqs.~(\ref{back1}) and (\ref{back2}), we obtain 
\be
L_N+\dot{\cal F}= \sum_{I=1}^{N-1} 
2X_I P^{(I)}_{X_I}\,.
\label{backd}
\ee
In Sec.~\ref{basec} we will show that the above k-essence 
description can accommodate non-relativistic matter and radiation
by choosing specific forms of $P^{(1)}(X_1)$ and $P^{(2)}(X_2)$.

In Ref.~\cite{LS} the conditions for eliminating derivatives higher than 
quadratic order were derived for the two-field action 
$S=\int d^4 x \sqrt{-g}\,L(N,K,{\cal S},{\cal R},{\cal Z},{\cal U},\phi_1,X_1;t)$. 
In this case, the higher-order spatial derivatives do not appear 
under the conditions (\ref{spacon1})-(\ref{spacon3}). 
The mixture of temporal and spatial derivatives higher 
than second order can be eliminated under the conditions 
$L_{KX_1}+2HL_{{\cal S}X_1}=0$ and 
$L_{{\cal R}X_1}+HL_{{\cal U}X_1}=0$ \cite{LS}.
For the separate Lagrangian 
$L(N,K,{\cal S},{\cal R},{\cal Z},{\cal U};t)+P^{(1)}(X_1)$, 
these two conditions are automatically satisfied. 
This is also the case for the action (\ref{Lagper}) of 
$N$ scalar fields. In the following we study the perturbations
for the action (\ref{Lagper}) with the Lagrangian (\ref{LH2}), 
in which case the higher-order temporal and spatial 
derivatives are absent.  

We now expand the action (\ref{Lagper}) up to 
second order in perturbations. In doing so, 
we express the three-dimensional metric $h_{ij}$ 
and the shift $N_i$ in the form \cite{Maldacena}
\ba
& &
h_{ij}=a^2(t)e^{2\zeta} \hat{h}_{ij}\,,\qquad
\hat{h}_{ij}=\delta_{ij}+\gamma_{ij}+\gamma_{il} 
\gamma_{lj}/2\,,\qquad 
{\rm det}\,\hat{h}=1\,,\\
& &N_i=\partial_{i} \psi \equiv \partial \psi/\partial x^i\,,
\ea
where $\zeta$ and $\psi$ are the scalar perturbations and 
$\gamma_{ij}$ is the tensor perturbation satisfying 
traceless and transverse conditions 
$\gamma_{ii}=\partial_{i}\gamma_{ij}=0$.

The second-order action for the tensor mode is the same 
as that derived in Refs.~\cite{Fedo,LS}:
\be
S_h^{(2)}=\int d^4 x \frac{a^3}{4} L_{\cal S} 
\left[ \dot{\gamma}^2_{ij}-c_t^2 
\frac{(\partial_k \gamma_{ij})^2}{a^2} \right]\,,
\label{Sh}
\ee
where the propagation speed $c_t$ is given by 
\be
c_t^2=\frac{{\cal E}}{L_{\cal S}}\,,\qquad
{\cal E} \equiv L_{\cal R}+\frac12 \dot{L}_{\cal U}
+\frac32 H L_{\cal U}\,.
\label{Edef}
\ee
The tensor ghosts and Laplacian instabilities are absent 
under the conditions 
\ba
L_{\cal S} &>& 0\,,
\label{noghostten} \\
{\cal E} &>& 0\,.
\label{insten}
\ea

For the scalar perturbations the second-order action can be
written in the form $S_s^{(2)}=\int d^4 x\,{\cal L}_2$, 
with the Lagrangian density
\ba
{\cal L}_2
&=& \delta \sqrt{h} [ (\dot{\cal F}+L_N) \delta N+{\cal E} 
\delta_1 {\cal R} ]
+a^3 [ (L_N+L_{NN}/2)\delta N^2+{\cal E} \delta_2 {\cal R}
+{\cal A} \delta K^2/2+{\cal B} \delta K \delta N+{\cal C} \delta K \delta_1
{\cal R} \nonumber \\
& & +({\cal D}+{\cal E}) \delta N \delta_1 {\cal R}+{\cal G} \delta_1 {\cal R}^2/2
+L_{\cal S} \delta K^{\mu}_{\nu} \delta K^{\nu}_{\mu}
+L_{\cal Z} \delta {\cal R}^{\mu}_{\nu} 
\delta {\cal R}^{\nu}_{\mu}]
+{\cal L}_2^{M}\,,
\label{L2}
\ea
where 
\ba
\hspace{-0.5cm}
& &
{\cal A} \equiv L_{KK}+4H L_{{\cal S}K}+4H^2 L_{\cal SS}\,,\qquad
{\cal B} \equiv L_{KN}+2HL_{{\cal S}N}\,,\qquad
{\cal C} \equiv L_{K{\cal R}}+2H L_{\cal SR}+L_{\cal U}/2
+H L_{K {\cal U}}+2H^2L_{\cal SU}\,,\nonumber \\
\hspace{-0.5cm}
& &
{\cal D} \equiv L_{N{\cal R}}-\dot{L_{\cal U}}/2+H L_{N{\cal U}}\,,
\qquad
{\cal G} \equiv L_{\cal RR}+2HL_{\cal RU}+H^2L_{\cal UU}\,.
\ea
The Lagrangian density ${\cal L}_2^{M}$ corresponds to the contribution 
coming from the matter fields $\phi_I$:
\be
{\cal L}_2^{M} \equiv \sum_{I=1}^{N-1} 
\left[ P^{(I)}_{X_I} \delta \sqrt{h}\,\delta_1 X_I+a^3 
\left( P^{(I)}_{X_I} \delta_2 X_I+P^{(I)}_{X_IX_I} \delta_1X_I^2/2
+P^{(I)}_{X_I}\delta N \delta_1 X_I \right)  \right]\,,
\ee
where the first-order and second-order contributions to $X_I$ are 
given, respectively, by 
\ba
& & \delta_1 X_I=2\dot{\phi}_I^2 \delta N-2\dot{\phi}_I \dot{\delta \phi}_I\,,\\
& & \delta_2 X_I=-\dot{\delta \phi}_I^2-3\dot{\phi}_I^2 \delta N^2
+4\dot{\phi}_I \dot{\delta \phi}_I \delta N
+\frac{2\dot{\phi}_I}{a^2} \partial_j \psi\partial_j \delta \phi_I
+\frac{1}{a^2}(\partial \delta \phi_I)^2 \,,
\ea
with $(\partial \delta \phi_I)^2 \equiv \partial_j \delta \phi_I 
\partial_j \delta \phi_I$ (the quantities with the same lower 
index $j$ are summed). On using Eq.~(\ref{backd}), 
one can eliminate some of the terms involving $\zeta$. 
Recall that the Lagrangian (\ref{LH2}) satisfies the 
relations (\ref{spacon1})-(\ref{spacon3}), i.e., 
${\cal A}+2L_{\cal S}=0$, 
${\cal C}=0$, and $4{\cal G}+3L_{\cal Z}=0$. 
On using the relations $\delta \sqrt{h}=3a^3 \zeta$, 
$\delta {\cal R}_{ij}=-(\delta_{ij}\partial^2 \zeta+\partial_i \partial_j \zeta)$, 
$\delta_1 {\cal R}=-4a^{-2} \partial^2 \zeta$, 
$\delta_2{\cal R}=-2a^{-2}[(\partial \zeta)^2-4\zeta \partial^2 \zeta]$, 
$\delta K^i_j=(\dot{\zeta}-H \delta N)\delta^i_j-\delta^{ik}
(\partial_k N_j+\partial_j N_k)/(2a^2)$, and 
$\delta K=3(\dot{\zeta}-H \delta N)-\partial^2 \psi/a^2$ \cite{Fedo,LS} 
with the notation $\partial^2 \zeta \equiv \partial_j \partial_j \zeta$, 
the Lagrangian density (\ref{L2}) can be expressed as
\ba
{\cal L}_2 &=& a^3 \biggl\{ \frac12 (2L_N+L_{NN}-6H{\cal W}
+12H^2L_{\cal S}) \delta N^2+
\left[ {\cal W} \left( 3\dot{\zeta}-\frac{\partial^2 \psi}{a^2} 
\right)-4 ({\cal D}+{\cal E}) \frac{\partial^2 \zeta}{a^2}
\right] \delta N +4L_{\cal S} \dot{\zeta} \frac{\partial^2 \psi}{a^2}
\nonumber \\
& &
-6L_{\cal S} \dot{\zeta}^2
+2{\cal E} \frac{(\partial \zeta)^2}{a^2} +\sum_{I=1}^{N-1} 
\biggl[ (2\dot{\phi}_I^2 P^{(I)}_{X_IX_I}
-P^{(I)}_{X_I}) (\dot{\phi}_I^2 \delta N^2-2\dot{\phi}_I
\dot{\delta \phi}_I \delta N+\dot{\delta \phi}_I^2)
-6\dot{\phi}_I P^{(I)}_{X_I} \zeta \dot{\delta \phi}_I \nonumber \\
& &
-2\dot{\phi}_I P^{(I)}_{X_I} \delta \phi_I 
\frac{\partial^2 \psi}{a^2}
+P^{(I)}_{X_I} \frac{(\partial \delta \phi_I)^2}{a^2} \biggr]
\biggr\}\,,
\label{L2f}
\ea
where 
\be
{\cal W} \equiv L_{KN}+2HL_{{\cal S}N}+4HL_{\cal S}\,.
\ee

Varying the Lagrangian density (\ref{L2f}) with 
respect to $\delta N$ and $\partial^2 \psi$, we obtain 
the Hamiltonian and momentum constraints
\ba
& & (2L_N+L_{NN}-6H{\cal W}+12H^2L_{\cal S}) \delta N
+ {\cal W} \left( 3\dot{\zeta}-\frac{\partial^2 \psi}{a^2} \right)
-4({\cal D}+{\cal E}) \frac{\partial^2 \zeta}{a^2} \nonumber \\
& &+\sum_{I=1}^{N-1} 2\dot{\phi}_I 
(P^{(I)}_{X_I}-2\dot{\phi}_I^2 P^{(I)}_{X_IX_I}) 
(\dot{\delta \phi}_I-\dot{\phi}_I \delta N)=0\,,\label{Hami}\\
& & {\cal W} \delta N-4L_{\cal S} \dot{\zeta}
+\sum_{I=1}^{N-1} 2\dot{\phi}_I P^{(I)}_{X_I} 
\delta \phi_I=0\,.
\label{momen}
\ea
Solving Eqs.~(\ref{Hami})-(\ref{momen}) for $\delta N$, $\partial^2 \psi$
and substituting the resulting relations into Eq.~(\ref{L2f}), the 
second-order Lagrangian density can be written in the form
\be
\mathcal{L}_{2}=a^{3}\left( \dot{\vec{\mathcal{X}}}^{t}{\bm K} 
\dot{\vec{\mathcal{X}}}-\frac{1}{a^{2}}\partial _{j}\vec{\mathcal{X}}^{t}{\bm G}
\partial _{j}{\vec{\mathcal{X}}}-\vec{\mathcal{X}}^{t}{\bm B} 
\dot{\vec{\mathcal{X}}}-\vec{\mathcal{X}}^{t}{\bm M} \vec{\mathcal{X}}\right) \,,
\label{L2mat}
\ee
where ${\bm K}$, ${\bm G}$, ${\bm B}$, ${\bm M}$ 
are $N \times N$ matrices, and 
the vector $\vec{\mathcal{X}}$ is composed from 
the dimensionless multiple fields, as 
\be
\vec{\mathcal{X}}^{t}=\left( \zeta, \delta \phi_1 /M_{\mathrm{pl}}, 
\cdots, \delta \phi_{N-1}/M_{\rm pl} \right) \,.
\ee
Here $M_{\rm pl}$ is the reduced Planck mass.

The two matrices ${\bm K}$ and ${\bm G}$ determine 
no-ghost conditions and the scalar propagation speeds. 
Their components are given by 
\ba
&& 
K_{11}=\frac{2L_{\cal S}}{{\cal W}^2} \left( g_2+
\frac{8L_{\cal S}}{M_{\rm pl}^2} \sum_{I=2}^{N}
\dot{\phi}_{I-1}^2 K_{II} \right)\,,\qquad
K_{II}=\left[ 2\dot{\phi}_{I-1}^2P^{(I-1)}_{X_{I-1}X_{I-1}}
-P^{(I-1)}_{X_{I-1}} \right] M_{\rm pl}^2\,, \nonumber \\
& &K_{1I}=K_{I1}=-\frac{4L_{\cal S}\dot{\phi}_{I-1}}
{M_{\rm pl}{\cal W}}K_{II}\,, \label{Kco}
\\
& &
G_{11}=-\frac12 \left( \dot{\cal C}_3+H{\cal C}_3+4{\cal E} \right)\,,\qquad
G_{II}=-P^{(I-1)}_{X_{I-1}}M_{\rm pl}^2\,,\qquad
G_{1I}=G_{I1}=\frac{{\cal C}_3\dot{\phi}_{I-1}}
{4L_{\cal S}M_{\rm pl}}G_{II}\,,  \label{Gco}
\ea
where $2 \le I \le N$ and other components are 0. 
The functions $g_2$ and ${\cal C}_3$ are defined by 
\ba
g_2 &\equiv&
4L_{\cal S} (2L_N+L_{NN})+3(L_{KN}+2HL_{{\cal S}N})^2\,,\\
{\cal C}_3 &\equiv& -\frac{16L_{\cal S}({\cal D}+{\cal E})}
{{\cal W}}\,.
\label{C3def}
\ea
For the derivation of $G_{11}$ we have used the property that 
the integral $\int d^4 x\,a{\cal C}_3 \dot{\zeta} \partial^2 \zeta$ 
reduces to $\int d^4 x\,(a/2)(\dot{\cal C}_3+H{\cal C}_3)
(\partial \zeta)^2$ up to a boundary term.

If the symmetric matrix ${\bm K}$ is positive definite, 
the scalar ghosts are absent. 
The necessary and sufficient conditions for the positivity of 
${\bm K}$ are that the determinants of principal submatrices 
of ${\bm K}$ are positive, i.e., 
\be
\frac{2 L_{\mathcal{S}}}{\mathcal{W}^2} \prod_{I=2}^{\ell} K_{II} 
\left(g_2+\frac{8 L_{\mathcal{S}}}{M_{\rm pl}^2} \sum_{J=\ell+1}^{N} 
\dot{\phi}_{J-1}^2K_{JJ}\right)>0 \qquad \quad 
(\ell=1,2,\cdots,N)\,,
\label{scagh1}
\ee
where we should understand that   
$\prod_{I=2}^{\ell} K_{II}=1$ for $\ell=1$ and 
$\sum_{J=\ell+1}^{N} \dot{\phi}_{J-1}^2K_{JJ}=0$ 
for $\ell=N$.
Under the tensor no-ghost condition (\ref{noghostten}), 
all the $N$ conditions (\ref{scagh1}) hold 
for $g_2>0$ and $K_{II}>0$ ($I=2,3,\cdots,N$). 
Hence the scalar ghost is absent for
\ba
&& 
g_2=4L_{\cal S} (2L_N+L_{NN})+3(L_{KN}
+2HL_{{\cal S}N})^2>0\,,
\label{noghost1} \\
&&
2\dot{\phi}_{I}^2P^{(I)}_{X_{I}X_{I}}-P^{(I)}_{X_{I}}>0\, \qquad
(I=1,2,\cdots,N-1).
\label{noghost2}
\ea

The dispersion relation following from the Lagrangian (\ref{L2mat}) 
in the limit of a large wave number $k$ with a frequency $\omega$
is given by 
\be
{\rm det} \left( \omega^2 {\bm K}-k^2{\bm G}/a^2 
\right)=0\,.
\label{cseq}
\ee
Introducing the scalar sound speed $c_s$ as 
$\omega^2=c_s^2\,k^2/a^2$, Eq.~(\ref{cseq}) reduces to 
\be
\prod_{I=1}^{N} \left( c_s^2\,K_{II}-G_{II} \right)
-\sum_{I=2}^{N} \left[  \left( c_s^2\,K_{1I}-G_{1I} \right)^2
\prod_{J \neq I,J \geq 2}^{N} \left( c_s^2\,K_{JJ}-G_{JJ} \right) 
\right]=0\,.
\label{cseq2}
\ee

For the theory described by the Lagrangian (\ref{LH2}), it follows that 
\be
{\cal D}+{\cal E}=B_4+B_{4N}-\frac12 H B_{5N}\,,\qquad
L_{\cal S}=-A_4-3HA_5\,.
\ee
We recall that in Horndeski theories the relation (\ref{Hocon}) holds, 
and hence ${\cal D}+{\cal E}=L_{\cal S}$. 
Then, the term ${\cal C}_3$ in Eq.~(\ref{C3def}) reads
\be
{\cal C}_{3{\rm H}}=-\frac{16L_{\cal S}^2}{{\cal W}}\,,
\label{C3H}
\ee
where the lower index ``H'' represents the values in 
Horndeski theories. 
Substituting the relation (\ref{C3H}) into Eq.~(\ref{Gco}) 
and using Eq.~(\ref{Kco}), the propagation 
speed $c_{s\rm H}$ in Horndeski theories satisfies 
\be
c_{s\rm H}^2 K_{1I}-G_{1I}=
-\frac{4L_{\cal S} \dot{\phi}_{I-1}}{M_{\rm pl}{\cal W}}
\left( c_{s\rm H}^2 K_{II} -G_{II} \right)\,.
\label{cHre}
\ee
Plugging Eq.~(\ref{cHre}) into Eq.~(\ref{cseq2}), 
we obtain the following algebraic equation
\be
\left[ c_{s{\rm H}}^2 K_{11}-G_{11}-\left( \frac{4L_{\cal S}}
{M_{\rm pl}{\cal W}} \right)^2 \sum_{I=2}^{N} 
\dot{\phi}_{I-1}^2 \left( c_{s{\rm H}}^2 K_{II}-G_{II} \right) 
\right] \prod_{I=2}^{N} \left( c_{s{\rm H}}^2 K_{II}-G_{II} 
\right)=0\,,
\label{cHeq}
\ee
whose solutions are given by 
\ba
& &
c_{s{\rm H}1}^2=\frac{G_{11} 
- [4L_{\cal S}/(M_{\rm pl}{\cal W})]^2  
\sum_{I=2}^{N} \dot{\phi}_{I-1}^2 G_{II}}{K_{11}
- [4L_{\cal S}/(M_{\rm pl}{\cal W})]^2 
\sum_{I=2}^{N} \dot{\phi}_{I-1}^2 K_{II}}=
\frac{{\cal W}^2}{2L_{\cal S}g_2} \left[ 
G_{11}+\frac{16L_{\cal S}^2}{{\cal W}^2} 
\sum_{I=2}^{N} \dot{\phi}_{I-1}^2 
P_{X_{I-1}}^{(I-1)} \right]\,,\label{cH1}\\
& & 
c_{s{\rm H}I}^2=\frac{G_{II}}{K_{II}}
=\frac{P_{X_{I-1}}^{(I-1)}}
{P_{X_{I-1}}^{(I-1)}-2\dot{\phi}_{I-1}^2
P_{X_{I-1}X_{I-1}}^{(I-1)}} \qquad (I=2,3,\cdots,N)\,.
\label{cH2}
\ea
The matter sound speed square (\ref{cH2}) coincides with that 
derived in Ref.~\cite{Garriga} in the context of single-field 
k-inflation. In Horndeski theories, each 
$c_{s{\rm H}I}$ ($I \geq 2$) is not affected by other scalar fields.
The presence of the matter fields $\phi_I$ gives rise to
modifications to the first propagation speed $c_{s{\rm H}1}$, which 
was already derived in Ref.~\cite{LS} for $N=2$.

In GLPV theories where the conditions (\ref{Hocon})
are not satisfied, we cannot write Eq.~(\ref{cseq2}) in the 
separate form like Eq.~(\ref{cHeq}).
On using the propagation speeds (\ref{cH1}) 
and (\ref{cH2}), Eq.~(\ref{cseq2}) can be written 
in the following form: 
\be
\prod _{I=1}^{N} \left( 
c_s^2 -c_{s{\rm H}I}^2 \right) 
=-\frac{8L_{\cal S}}{g_2} 
\left( \frac{{\cal C}_3 {\cal W}}{16L_{\cal S}^2}+1 \right)
\sum_{I=2}^{N} \biggl[ \dot{\phi}_{I-1}^2 
P_{X_{I-1}}^{(I-1)} \left\{ 2c_s^2+c_{s{\rm H}I}^2 
\left( \frac{{\cal C}_3 {\cal W}}{16L_{\cal S}^2}-1 
\right) \right\} \prod_{J \neq I, J\geq 2}^{N}
(c_s^2-c_{s{\rm H}J}^2) \biggr]\,,
\label{ceq}
\ee
where, for $N=2$, $\prod_{J \neq I, J\geq 2}^{N}
(c_s^2-c_{s{\rm H}J}^2)=1$.
Since ${\cal C}_3 \neq -16L_{\cal S}^2/{\cal W}$ in GLPV theories, 
the right hand side of Eq.~(\ref{ceq}) does not vanish. 
Hence $c_s^2$ differs from the value $c_{s{\rm H}I}^2$.
This means that not only the propagation speed $c_{s{\rm H}1}$
but also the matter sound speeds $c_{s{\rm H}I}$ ($I \geq 2$) 
are affected by the presence of other scalar fields. 
When $N=2$, Eq.~(\ref{ceq}) reduces to 
\be
\left( c_s^2-c_{s{\rm H}1}^2 \right) \left( c_s^2-c_{s{\rm H}2}^2 \right)
=-\frac{8L_{\cal S}}{g_2} \left( \frac{{\cal C}_3 {\cal W}}
{16L_{\cal S}^2} +1\right) \dot{\phi}_1^2 P_{X_1}^{(1)} 
\left[ 2c_s^2+c_{s{\rm H}2}^2 
\left( \frac{{\cal C}_3 {\cal W}}{16L_{\cal S}^2}-1 
\right) \right] \,,
\label{ceqN=2}
\ee
which agrees with Eq.~(22) of 
Ref.~\cite{Gleyzes}.\footnote{In Ref.~\cite{Gleyzes} 
the definition of the first propagation speed square is given by 
\be
\tilde{c}_{s{\rm H1}}^2=\frac{{\cal W}^2}{2L_{\cal S}g_2} 
\left[ G_{11}+\left( \frac{{\cal C}_3}{4L_{\cal S}} \right)^2
\dot{\phi}_1^2P_{X_1}^{(1)} \right]\,,\nonumber
\ee
whereas the definition of $c_{s{\rm H2}}^2$ is the same as ours.
In the Horndeski limit ${\cal C}_3 \to -16L_{\cal S}^2/{\cal W}$, 
$c_{s{\rm H1}}^2$ is identical to $\tilde{c}_{s{\rm H1}}^2$.}

Let us consider the case in which the deviation of 
${\cal C}_3$ from the value $-16L_{\cal S}^2/{\cal W}$ 
is small, i.e., 
\be
{\cal C}_3=-\frac{16L_{\cal S}^2}{{\cal W}} 
\left( 1+\delta {\cal C}_3 \right)\,,\qquad
|\delta {\cal C}_3| \ll 1\,.
\label{delC3def}
\ee
Under this approximation we write the two solutions 
for $c_s^2$ in Eq.~(\ref{ceq}), as 
\ba
c_{s1}^2 &=& 
c_{s{\rm H1}}^2+\delta c_{s1}^2\,,\label{cs1so} \\
c_{sI}^2 &=&
c_{s{\rm H}I}^2+\delta c_{sI}^2 
\qquad (I=2,3,\cdots,N)\,.
\label{cs2so} 
\ea
Substituting Eq.~(\ref{cs1so}) into Eq.~(\ref{ceq}), we obtain 
\be
\delta c_{s1}^2 \simeq \sum_{I=2}^{N} \xi_{I-1}\,\delta {\cal C}_3\,,
\label{delcH1}
\ee
where 
\be
\xi_I \equiv \frac{16L_{\cal S} \dot{\phi}_I^2P_{X_I}^{(I)}}{g_2}\,.
\label{xidef}
\ee
When we substitute the solution (\ref{cs2so}) into Eq.~(\ref{ceq}),
we employ the approximation $|\delta c_{sI}^2| \ll c_{s{\rm H}I}^2$, 
whose validity should be checked after deriving 
the solution of $\delta c_{sI}^2$. 
It then follows that 
\be
\delta c_{sI}^2 \simeq
-\frac{c_{s{\rm H}I}^2}{2(c_{s{\rm H}I}^2-c_{s{\rm H1}}^2
-\delta c_{s1}^2)} \xi_{I-1} \delta {\cal C}_3^2
\qquad (I=2,3,\cdots,N).
\label{delcH2}
\ee

If the quantities $|\xi_{I-1}|$ ($I \geq 2$) are much larger than 1, 
it is possible to have $|\delta c_{s1}^2|$ of the order of 1 
even for $|\delta {\cal C}_3| \ll 1$. 
In fact, this happens for the cosmology of the covariantized 
Galileon studied in Sec.~\ref{Galisec}.
On the other hand, $\delta c_{sI}^2$ ($I \geq 2$) 
contains an additional suppression factor $\delta {\cal C}_3$. 
In the cosmological epoch where the field $\phi_{I-1}$ ($I \geq 2$)
dominates the energy density of the Universe, we have 
$\delta c_{s1}^2 \simeq \xi_{I-1}\,\delta {\cal C}_3$ from 
Eq.~(\ref{delcH1}).
Provided that the terms $|\delta c_{s1}^2|$ and 
$|c_{s{\rm H}I}^2-c_{s{\rm H1}}^2|$ 
are at most of the order of 1, it follows that 
$|\delta c_{sI}^2| \ll c_{s{\rm H}I}^2$.
This discussion implies that the deviation from Horndeski theories 
may potentially lead to a considerable modification to $c_{s\rm H1}^2$, 
but the modification to $c_{s\rm HI}^2$ ($I \geq 2$) should be suppressed. 
In Sec.~\ref{prosec} we shall study this issue for 
concrete models of dark energy.


\section{Application to Galileon theories}
\label{Galisec} 

The covariant Galileon advocated in Ref.~\cite{Galileon} belongs to a class 
of the Horndeski Lagrangian (\ref{LH}) with the functions
\be
G_2=\frac{c_2}{2}Y\,,\qquad
G_3=\frac{c_3}{2M^3}Y\,,\qquad
G_4=\frac{M_{\rm pl}^2}{2}-\frac{c_4}{4M^6}Y^2\,,\qquad
G_5=\frac{3c_5}{4M^9}Y^2\,,
\label{Gali1}
\ee
where $c_{2,3,4,5}$ are dimensionless constants and $M$ is a constant 
having a dimension of mass. In this case the auxiliary functions 
$F_3$ and $F_5$ can be chosen as 
$F_3=c_3Y/(6M^3)$ and $F_5=3c_5Y^2/(5M^9)$, respectively. 
Then, the covariant Galileon [dubbed Model (A)] corresponds to the 
Lagrangian (\ref{LH2}) with the functions
\ba
{\rm Model~(A)}:& & 
A_2=\frac{c_2}{2}Y\,,\qquad 
A_3=\frac{c_3}{3M^3} (-Y)^{3/2}\,,\qquad
A_4=-\frac{M_{\rm pl}^2}{2}-\frac{3c_4}{4M^6}Y^2\,,\qquad
A_5=\frac{c_5}{2M^9} (-Y)^{5/2}\,,\nonumber \\
& &
B_4=\frac{M_{\rm pl}^2}{2}-\frac{c_4}{4M^6}Y^2\,,\qquad
B_5=-\frac{3c_5}{5M^9}(-Y)^{5/2}\,.
\label{cova1}
\ea

The Lagrangian of the original Galileon \cite{Nicolis} 
was constructed such that 
the field equations of motion satisfy the Galilean symmetry 
$\partial_{\mu} \phi \to \partial_{\mu}\phi+b_{\mu}$ 
in Minkowski space-time. 
In curved space-time, the covariantized version of 
the Minkowski Galileon follows by replacing partial 
derivatives of the field with covariant derivatives. 
Although this process generally gives rise to derivatives 
higher than second order, the equations of motion for the
covariantized Galileon remain of second order 
on the isotropic cosmological background. 
Taking into account the Einstein-Hilbert term 
$(M_{\rm pl}^2/2)R$, the Lagrangian of the 
covariantized Galileon [dubbed Model (B)] is given by 
\ba
{\rm Model~(B)}:& & 
A_2=\frac{c_2}{2}Y\,,\qquad 
A_3=\frac{c_3}{3M^3} (-Y)^{3/2}\,,\qquad
A_4=-\frac{M_{\rm pl}^2}{2}-\frac{3c_4}{4M^6}Y^2\,,\qquad
A_5=\frac{c_5}{2M^9} (-Y)^{5/2}\,,
\nonumber \\
& &
B_4=\frac{M_{\rm pl}^2}{2}\,,\qquad
B_5=0\,.
\label{cova2}
\ea
The additional terms $-c_4Y^2/(4M^6)$ and $-3c_5(-Y)^{5/2}/(5M^9)$ 
appearing in the terms $B_4$ and $B_5$ of the covariant Galileon 
Lagrangian (\ref{cova1}) correspond to the gravitational 
counter terms that eliminate 
derivatives higher than second order in general space-time.

\subsection{Background cosmology}
\label{basec}

Even though the coefficients $B_4$ and $B_5$ in Model (B) 
are different from those in Model (A), the 
coefficients $A_i$ ($i=2,3,4,5$) are the same in both cases.
Hence the background cosmological dynamics in Model (B) 
are exactly the same as those in Model (A).
Substituting the functions $A_i$ of Eqs.~(\ref{cova1}) and 
(\ref{cova2}) into Eqs.~(\ref{back1d}) and (\ref{back2d}), 
we obtain the equations of motion 
\ba
& &
3M_{\rm pl}^2 H^2=\rho_{\rm DE}+\rho_M\,,
\label{backeq1} \\
& &
3M_{\rm pl}^2 H^2+2M_{\rm pl}^2 \dot{H}
=-P_{\rm DE}-P_M\,,
\label{backeq2}
\ea
where the energy density $\rho_{\rm DE}$ and the pressure 
$P_{\rm DE}$ of the ``dark'' component are given by 
\ba
\rho_{\rm DE} &=&
-\frac12 c_2 \dot{\chi}^2+\frac{3c_3H \dot{\chi}^3}{M^3}
-\frac{45c_4H^2 \dot{\chi}^4}{2M^6}+\frac{21c_5H^3\dot{\chi}^5}{M^9}\,,\\
P_{\rm DE} &=& -\frac12 c_2 \dot{\chi}^2-\frac{c_3\dot{\chi}^2 \ddot{\chi}}
{M^3}+\frac{3c_4 \dot{\chi}^3}{2M^6}
\left[ 8H \ddot{\chi}+(3H^2+2\dot{H})\dot{\chi} \right]
-\frac{3c_5H \dot{\chi}^4}{M^9}
\left[ 5H \ddot{\chi} +2(H^2+\dot{H}) \dot{\chi} \right]\,.
\ea
Equations (\ref{backeq1}) and (\ref{backeq2}) coincide with those derived 
in Refs.~\cite{DT10PRL,DT10PRD} for the covariant Galileon. 
The dark energy equation of state is defined by 
$w_{\rm DE} \equiv P_{\rm DE}/\rho_{\rm DE}$.

For the matter component labelled by the lower 
index ``$M$'' in Eqs.~(\ref{backeq1})-(\ref{backeq2}), 
we take into account radiation and non-relativistic matter.
The perfect fluids of radiation and non-relativistic 
matter can be modeled by two scalar fields $\phi_1$ 
and $\phi_2$, respectively, with the Lagrangians 
\ba
&&
P^{(1)} (X_1)=
b_1 X_1^2\,,
\label{P1}\\
&& 
P^{(2)} (X_2)=
b_2 (X_2-X_0)^2\,,
\label{P2}
\ea
where $b_1$, $b_2$, and $X_0$ are constants. 
If we add a constant term $\Lambda$ to the Lagrangian (\ref{P2}), 
this corresponds to the unified model of dark matter and dark energy 
proposed by Scherrer \cite{Scherrer}.

{}From Eq.~(\ref{rhowI}) the energy density and the equation 
of state of radiation are given, respectively, by  
$\rho_r=3b_1 X_1^2$ and $w_r=1/3$. 
The no-ghost condition (\ref{noghost2}) is satisfied for $b_1>0$.
In Horndeski theories including Model (A), 
Eq.~(\ref{cH2}) shows that the sound speed
square of radiation is given by 
\be
c_{s{\rm H2}}^2=\frac13\,.
\label{csrad}
\ee
In Model (B), the radiation sound speed square $c_{s2}^2$ 
deviates from $c_{s{\rm H2}}^2$ 
with the difference estimated by Eq.~(\ref{delcH2}). 

The energy density and the equation of state of non-relativistic 
matter following from Eq.~(\ref{P2}) are given, respectively, by 
\be
\rho_m=b_2 (X_2-X_0) (3X_2+X_0)\,,\qquad
w_m=\frac{X_2-X_0}{3X_2+X_0}\,.
\ee
Provided that $X_2$ is close to $X_0$, the field $\phi_2$ 
behaves as non-relativistic matter with $w_m \simeq 0$. 
Then, the no-ghost condition (\ref{noghost2}) is satisfied 
for $b_2>0$.
{}From the continuity equation 
$\dot{\rho}_m+3H(1+w_m) \rho_m=0$, 
we obtain the dependence $\rho_m \propto a^{-3}$ for 
$\epsilon \equiv (X_2-X_0)/X_0 \ll 1$ and hence 
$\epsilon \propto w_m \propto a^{-3}$. 
The matter energy density can be expressed 
as $\rho_m =16b_2 X_0^2w_m/(1-3w_m)^2$. 
In Horndeski theories, the sound speed 
square of non-relativistic matter reads
\be
c_{s{\rm H3}}^2=\frac{X_2-X_0}{3X_2-X_0}
=\frac{2w_m}{1+3w_m}\,,
\label{cH3}
\ee
which is much smaller than 1 for $w_m \ll 1$.
The matter sound speed square $c_{s3}^2$ in Model (B)
is subject to change relative to $c_{s{\rm H3}}^2$ 
given above. 
The presence of an additional pressure affects the 
gravitational growth of matter perturbations.
For the successful structure formation, we require that 
$c_{s3}^2 \ll 1$ during the matter-dominated epoch. 

The background cosmology based on Eqs.~(\ref{backeq1}) 
and (\ref{backeq2}) has been studied in detail 
in Refs.~\cite{DT10PRL,DT10PRD}. 
In what follows we shall briefly review the background dynamics and then 
study how the two Galileon theories can be distinguished from each other 
at the level of perturbations.
A de Sitter solution ($H=H_{\rm dS}={\rm constant}$) responsible for 
the late-time cosmic acceleration can be realized for
a constant field velocity $\dot{\chi}_{\rm dS}$. 
Normalizing the mass $M$ as $M^3=M_{\rm pl}H_{\rm dS}^2$ and 
defining the dimensionless variable
$x_{\rm dS} \equiv \dot{\chi}_{\rm dS}/(H_{\rm dS}M_{\rm pl})$, 
the coefficients $c_2$ and $c_3$ are related with the quantities 
$\alpha \equiv c_4 x_{\rm dS}^4$ and 
$\beta \equiv c_5 x_{\rm dS}^5$, as
\be
c_2 x_{\rm dS}^2=6+9\alpha-12\beta\,,\qquad
c_3 x_{\rm dS}^3=2+9\alpha-9\beta\,. 
\label{dscondition}
\ee
We also introduce the following dimensionless variables:
\be
r_1 \equiv \frac{\dot{\chi}_{\rm dS}H_{\rm dS}}{\dot{\chi}H}\,,
\qquad
r_2 \equiv \frac{H}{H_{\rm dS}} \left( \frac{\dot{\chi}}
{\dot{\chi}_{\rm dS}} \right)^5\,,
\label{r1r2}
\ee
which are normalized as $r_1=r_2=1$ at the de Sitter fixed point.
The Friedmann equation (\ref{backeq1}) can be written in the form
\be
\Omega_m =1-\Omega_r-\Omega_{\rm DE}\,,
\label{Omem}
\ee
where 
$\Omega_m \equiv \rho_m/(3M_{\rm pl}^2 H^2)$, 
$\Omega_r \equiv \rho_r/(3M_{\rm pl}^2 H^2)$, and 
\be
\Omega_{\rm DE} \equiv \frac{\rho_{\rm DE}}
{3M_{\rm pl}^2 H^2}
= -\frac12 (2+3\alpha-4\beta)r_1^3r_2
+(2+9\alpha-9\beta)r_1^2r_2
-\frac{15}{2}\alpha r_1 r_2+7\beta r_2\,.
\label{OmeDE}
\ee
The autonomous equations of motion for the variables 
$r_1$, $r_2$, and $\Omega_r$ are presented 
in Appendix (see also Ref.~\cite{DT10PRD} for detail).
The variation of the Hubble parameter is known by 
$H'/H=-5r_1'/(4r_1)-r_2'/(4r_2)$, where 
a prime represents the derivative with respect to $\ln a$.

There exists a so-called tracker solution characterized by 
$r_1=1$, along which the field velocity evolves as 
$\dot{\chi} \propto H^{-1}$ \cite{DT10PRL,DT10PRD}. 
Along the tracker, the variable $r_2$ grows as 
$r_2'=2r_2(3-3r_2+\Omega_r)/(1+r_2)$ from the regime
$r_2 \ll 1$ to the de Sitter fixed point characterized by 
$r_2=1$ and $\Omega_r=0$. 
The dark energy equation of state on the tracker is given by 
\be
w_{\rm DE}=-\frac{\Omega_r+6}{3(1+r_2)}\,,
\label{wdetra}
\ee
which evolves as $w_{\rm DE}=-7/3 \to -2 \to -1$ during the 
cosmological sequence of radiation 
($\Omega_r \simeq 1$, $r_2 \ll 1$), 
matter ($\Omega_r \ll 1$, $r_2 \ll 1$), and de Sitter 
($\Omega_r \ll 1$, $r_2=1$) epochs.
However, the tracker equation of state (\ref{wdetra}) is in 
tension with the joint data analysis of Sn Ia, CMB, and 
BAO because of the large deviation of $w_{\rm DE}$ from 
$-1$ during the matter era \cite{Nesseris}. 

The solutions that approach the tracker at late times 
can be consistent with the observational data.
In this case, the variable $r_1$ is much smaller than 1 
during the early stage of the cosmological evolution.
In the regime $r_1 \ll 1$, the variables $r_1$ and $r_2$ 
approximately obey the differential equations
\be
r_1' \simeq \frac{9+\Omega_r+21 \beta r_2}
{8+21 \beta r_2} r_1\,,\qquad
r_2' \simeq\frac{3+11\Omega_r-21\beta r_2}
{8+21\beta r_2}r_2\,.
\ee
Provided that $|\beta r_2| \ll 1$, we obtain the solutions 
$r_1 \propto a^{5/4}$, $r_2 \propto a^{7/4}$
during the radiation era and $r_1 \propto a^{9/8}$, 
$r_2 \propto a^{3/8}$ during the matter era. 
When $r_1 \ll 1$, the dark energy equation of state
is given by  
\be
w_{\rm DE} \simeq -\frac{1+\Omega_r}{8+21\beta r_2}\,.
\ee
In the regime $|\beta r_2| \ll 1$, $w_{\rm DE}$ evolves from 
the value $-1/4$ (radiation era) to the value $-1/8$ (matter era).
Once $r_1$ approaches 1, the solutions enter the tracking
regime characterized by the equation of state (\ref{wdetra}).
Provided that the approach to the tracker occurs at low 
redshifts, $w_{\rm DE}$ takes a minimum value larger 
than $-1.3$ and then it approaches the de Sitter value 
$-1$ in the asymptotic future. 
Such late-time tracking solutions are consistent with the combined 
data analysis of Sn Ia, CMB, and BAO \cite{Nesseris}. 

\subsection{No-ghost conditions}

Let us discuss no-ghost conditions for tensor and scalar perturbations 
in Models (A) and (B).
{}From Eq.~(\ref{noghostten}) the tensor ghost is absent for 
$L_{\cal S}=-A_4-3HA_5>0$, whose condition is the same in 
both Models (A) and (B).
This property also holds for no-ghost conditions 
of the scalar mode, because Eq.~(\ref{scagh1}) does not 
involve the functions $B_4, B_5$ and their derivatives. 
We recall that, for the matter fields characterized by the 
Lagrangians (\ref{P1}) and (\ref{P2}), the conditions 
(\ref{noghost2}) are satisfied for $b_1>0$ and 
$b_2>0$, respectively.

In the following, let us consider the case in which the sign of 
$\dot{\chi}$ does not change during the cosmic expansion 
history, i.e., $r_1>0$ and $r_2>0$. 
On using the variables $r_1$ and $r_2$, the no-ghost conditions 
(\ref{noghostten}) and (\ref{noghost1}) for tensor and scalar 
modes are given, respectively, by 
\ba
\hspace{-0.3cm}
L_{\cal S}&=&
[2+3r_2 (\alpha r_1-2\beta)]M_{\rm pl}^2/4>0\,,
\label{ghostcon1} \\
\hspace{-0.3cm}
g_2 &=& 3M_{\rm pl}^4 H_{\rm dS}^2 \sqrt{r_2/r_1^5} \nonumber \\
\hspace{-0.3cm}
& & \times
\{(72\alpha^2+81\beta^2-150\alpha \beta+30\alpha-36 \beta+4)r_1^4 r_2
+[8\beta-6\alpha-4-(162 \alpha^2+24\beta^2-180\alpha \beta
+36\alpha-12\beta)r_2]r_1^3 \nonumber \\
\hspace{-0.3cm}
& &~~~+[36 \alpha-36\beta+8+(90\alpha^2-162\beta^2
+162\alpha \beta+36\beta)r_2]r_1^2
-12\alpha (3+16\beta r_2)r_1+105\beta^2 r_2+40\beta \}>0\,.
\label{ghostcon2}
\ea
In the regime $r_1 \ll 1$ and $r_2 \ll 1$ the condition (\ref{ghostcon1}) is 
satisfied, while another condition (\ref{ghostcon2}) translates to 
\be
\beta>0\,.
\label{noghost0}
\ee
In order to satisfy Eqs.~(\ref{ghostcon1}) and (\ref{ghostcon2}) in the 
tracking regime ($r_1=1$ and $0<r_2 \le 1$), we require that 
\be
-2<3(\alpha-2\beta)<2\,.
\label{noghost}
\ee
The conditions (\ref{noghost0}) and (\ref{noghost}) need to obey 
for avoiding tensor and scalar ghosts.

\subsection{Tensor propagation speeds}

Since ${\cal E}=B_4+\dot{B}_5/2$ for the Lagrangian (\ref{LH2}), 
the tensor propagation speed square $c_t^2={\cal E}/L_{\cal S}$
is different between the two Galileon theories.
For Model (A) it is given by \cite{DT10PRL,DT10PRD}
\be
c_{t}^2=\frac{2r_1 (2-\alpha r_1 r_2)-3\beta (r_1r_2'+r_2r_1')}
{2r_1[2+3r_2 (\alpha r_1-2\beta)]} \qquad \quad [{\rm Model~(A)}],
\label{ctmodelA}
\ee
which is close to 1 for $r_2 \ll 1$.
At the de Sitter fixed point we have $c_{t}^2=(2-\alpha)/(2+3\alpha-6\beta)$, 
so we require $\alpha<2$ to avoid the 
Laplacian instability of tensor perturbations under the condition (\ref{noghost}).
During the transition from the regime $r_1=1, r_2 \ll 1$ to 
the regime $r_1=1, r_2=1$, it happens that $c_t^2$ has a minimum.
Imposing that $c_{t}^2>0$ at the minimum, it follows that 
$\alpha<12 \sqrt{\beta}-9\beta-2$ \cite{DT10PRL,DT10PRD}.

For Model (B) we have ${\cal E}=M_{\rm pl}^2/2$, 
so the tensor propagation speed square is simply given by 
\be
c_t^2=\frac{2}{2+3r_2 (\alpha r_1-2\beta)}
 \qquad \quad [{\rm Model~(B)}]\,.
\ee
Under the no-ghost condition (\ref{ghostcon1}), $c_t^2$ is positive.
As long as the tensor perturbation is concerned, the viable parameter 
space of Model (B) is not restrictive compared to 
that of Model (A).

\subsection{Scalar propagation speeds}
\label{prosec}

\subsubsection{{\rm Model (A)}}

The covariant Galileon model (A) belongs to a class of Horndeski theories, 
so the three scalar propagation speed squares $c_s^2$
follow from Eqs.~(\ref{cH1})-(\ref{cH2}) with $I=2,3$. 
Among them the sound speed squares 
$c_{s2}^2$ and $c_{s3}^2$ of radiation and non-relativistic 
matter are given, respectively,  
by Eqs.~(\ref{csrad}) and (\ref{cH3}).
On using the relation 
\be
\dot{\phi}_I^2 P^{(I)}_{X_I}=-\frac12 (\rho^{(I)}+P^{(I)})
=-\frac32 M_{\rm pl}^2 H^2(1+w_{(I)})\Omega_{(I)}\,,
\ee
where $w_{(I)}$ and $\Omega_{(I)}$ are the equation of state 
and the density parameter of the field $\phi_I$, 
the first propagation speed square $c_{s1}^2$ 
follows from Eq.~(\ref{cH1}), as
\be
c_{s1}^2=-\frac{{\cal W}^2 M_{\rm pl}^2}{4L_{\cal S}g_2} 
\left[ \left( 1-\frac{H'}{H} \right) \tilde{C}_{3{\rm H}}+\tilde{C}_{3{\rm H}}'+
\frac{4{\cal E}_{\rm H}}{M_{\rm pl}^2}+\frac{48L_{\cal S}^2H^2}
{{\cal W}^2} \left\{ \Omega_r (1+w_r)+\Omega_m (1+w_m) 
\right\} \right]\,,
\label{cH1co}
\ee
where $\tilde{C}_{3{\rm H}}=H {\cal C}_{3{\rm H}}/M_{\rm pl}^2$, 
${\cal E}_{\rm H}=[2r_1 (2-\alpha r_1 r_2)-3\beta (r_1r_2'+r_2r_1')]
M_{\rm pl}^2/(8r_1)$, and 
\be
{\cal W}=M_{\rm pl}^2 H_{\rm dS}(r_1^5r_2)^{-1/4}
\left[2-21\beta r_2+15\alpha r_1r_2 -(2+9\alpha-9\beta)r_1^2 r_2
\right]\,.
\ee
Note that the matter density parameter $\Omega_m$ can 
be eliminated by using the relation (\ref{Omem}). 
The evolution of $c_{s1}^2$ in three asymptotic 
regimes is given by \cite{DT10PRL,DT10PRD}:
\ba
c_{s1}^2=
\left\{
\begin{aligned}
& \frac{1}{40} (\Omega_r+1) \qquad \qquad 
\qquad \qquad \qquad \qquad\,\,\,
[{\rm (i)}~r_1 \ll 1,~r_2 \ll 1]\,,\\
& \frac{8+10\alpha-9\beta+\Omega_r (2+3\alpha-3\beta)}
{3(2-3\alpha+6\beta)} \qquad \quad
[{\rm (ii)}~r_1=1,~r_2 \ll 1]\,, \\
& \frac{(\alpha-2\beta)(4+15\alpha^2-48\alpha \beta
+36\beta^2)}{2(2+3\alpha-6\beta)(2-3\alpha+6\beta)}
\qquad 
[{\rm (iii)}~r_1=1,~r_2=1]\,.
\end{aligned}
\right.
\label{csA}
\ea
In the regime (i) we have $c_{s1}^2=1/20$ and $1/40$ during 
the radiation and matter eras, respectively, so there is no 
Laplacian instability. If the solutions enter the tracking regime (ii)
during the radiation era, we require the stability condition 
$10+13\alpha-12\beta>0$. 
Taking into account the condition (\ref{noghost}), 
the de Sitter fixed point (iii) is stable for $\alpha>2\beta$. 
The theoretically viable parameter space is shown in figure 1 
of Ref.~\cite{DT10PRL}. The evolution of matter density perturbations 
and observational constraints on the covariant Galileon from 
large-scale structures have been studied in Refs.~\cite{Galima}.

\subsubsection{{\rm Model (B)}}

In the case of Model (B), we need to solve the coupled 
equation (\ref{cseq2}) for $N=3$, i.e., 
\be
\left( c_s^2K_{11}-G_{11} \right) \left( c_s^2K_{22}-G_{22} \right)
\left( c_s^2K_{33}-G_{33} \right)
-\left( c_s^2K_{12}-G_{12} \right)^2 \left( c_s^2K_{33}-G_{33} \right)
-\left( c_s^2K_{13}-G_{13} \right)^2 \left( c_s^2K_{22}-G_{22} \right)
=0\,.
\ee
The solutions to this third-order equation 
for $c_s^2$ are given by 
\be
c_s^2=-\frac{a_2}{3a_1}+u_{+}+u_{-}\,,\quad
-\frac{a_2}{3a_1}+u_{+}\omega
+u_{-}\omega^2\,,
\quad 
-\frac{a_2}{3a_1}+u_{+}\omega^2
+u_{-}\omega\,,
\label{csol}
\ee
where $\omega=-(1+\sqrt{3}\,i)/2$, 
$u_{\pm}=[(-q \pm \sqrt{q^2+4p^3/27})/2]^{1/3}$, 
$p=a_3/a_1-a_2^2/(3a_1^2)$, 
$q=2a_2^3/(27a_1^3)-a_2a_3/(3a_1^2)+a_4/a_1$, and 
\ba
a_1 &=& K_{11}K_{22}K_{33}-K_{12}^2 K_{33}
-K_{13}^2 K_{22}\,,\\
a_2 &=& K_{12}^2 G_{33}+K_{13}^2 G_{22}
-K_{11}K_{22}G_{33}-K_{11}G_{22}K_{33}
-G_{11}K_{22}K_{33}+2K_{12}G_{12}K_{33}
+2K_{13}G_{13}K_{22}\,,\\
a_3 &=& K_{11}G_{22}G_{33}+G_{11}K_{22}G_{33}
+G_{11}G_{22}K_{33}-G_{12}^2K_{33}-G_{13}^2K_{22}
-2K_{12}G_{12}G_{33}-2K_{13}G_{13}G_{22}\,,\\
a_4 &=& G_{12}^2 G_{33}+G_{13}^2G_{22}
-G_{11}G_{22}G_{33}\,.
\ea
When $q^2+4p^3/27<0$, all the solutions (\ref{csol}) are real.

One of the solutions $c_{s1}^2$ in Eq.~(\ref{csol}) is associated with 
the propagation speed square of the dark energy field $\chi$.
In three asymptotic regimes it is given by 
\ba
c_{s1}^2=\left\{
\begin{aligned}
& \frac{1}{40} (3\Omega_r-1) \qquad \qquad 
\qquad \qquad \qquad \qquad \qquad
[{\rm (i)}~r_1 \ll 1,~r_2 \ll 1]\,,\\
& \frac{16-15(\alpha-2\beta)
+\Omega_r (4-3\alpha+6\beta)}
{6(2-3\alpha+6\beta)} \qquad \quad~~\,
[{\rm (ii)}~r_1=1,~r_2 \ll 1]\,, \\
& \frac{\alpha-2\beta}{2+3\alpha-6\beta}
\qquad \qquad 
\qquad \qquad \qquad \qquad \qquad
[{\rm (iii)}~r_1=1,~r_2=1]\,.
\label{csse}
\end{aligned}
\right.
\ea
Under the no-ghost condition (\ref{noghost}), the propagation 
speed square $c_{s1}^2$ in the regime (ii) is positive.
The de Sitter fixed point (iii) is stable for 
\be
\alpha>2\beta\,.
\ee
In the regime (i) we have $c_{s1}^2=1/20$ for $\Omega_r=1$, 
but $c_{s1}^2=-1/40$ for $\Omega_r=0$. 
This means that the perturbations are plagued by 
short-scale Laplacian instabilities during the matter era
for late-time tracking solutions.
As long as the solutions approach the tracker by the end of 
the radiation era, it is possible to avoid the Laplacian instability 
of scalar perturbations.
We recall however that only the background trajectories 
approaching the tracker around the end of the matter era are
consistent with the joint data analysis of Sn Ia, CMB, 
and BAO \cite{Nesseris}. 
Then the solutions need to be in the regime (i) during 
most of the matter era, in which case the Laplacian instability
cannot be avoided.

In the regime (i), the quantity $c_{s{\rm H}1}$ defined by 
Eq.~(\ref{cH1}) evolves as
\be
c_{s{\rm H}1}^2=\frac{1}{40} \left( 7\Omega_r+11 \right)\,,
\label{csHB}
\ee
which is positive.
The difference between (\ref{csHB}) and $c_s^2=(3\Omega_r-1)/40$ 
in Eq.~(\ref{csse}) should be induced from the term $\delta {\cal C}_3$ 
in Eq.~(\ref{delC3def}).
This term can be expressed as
\be
\delta {\cal C}_3=-\frac{3r_2(\alpha r_1-2\beta)}
{2+3r_2 (\alpha r_1-2\beta)}\,,
\ee
which means that $|\delta {\cal C}_3| \ll 1$ in the regimes (i) and (ii). 
For radiation, the quantity $\xi_1$ defined by Eq.~(\ref{xidef})
evolves as $\xi_1=-2/(15\beta r_2)$ in the regime (i) and 
hence $|\xi_1| \gg 1$.
{}From Eq.~(\ref{delcH1}) we then have 
$\delta c_{s1}^2=-2/5$ during the 
radiation era, so that $c_{s1}^2=c_{s{\rm H}1}^2
+\delta c_{s1}^2=1/20$.
For non-relativistic matter, the evolution of the 
quantity $\xi_2$ is given by $\xi_2=-1/(10\beta r_2)$ 
and hence $\delta c_{s1}^2=-3/10$
during the regime (i) of the matter era. 
Hence we obtain the negative propagation speed square
$c_{s1}^2=c_{s{\rm H}1}^2+\delta c_{s1}^2=-1/40$.
Interestingly, even if the difference between ${\cal C}_3$ 
and ${\cal C}_{3{\rm H}}$ is small in the regime (i), 
the modification to $c_{s{\rm H}1}^2$ 
cannot be negligible. 
{}From Eq.~(\ref{delcH2}) the corrections $\delta c_{sI}^2$ ($I=2,3$) 
to $c_{s{\rm H}2}^2$ and $c_{s{\rm H}3}^2$ 
of radiation and non-relativistic matter are suppressed relative to 
$\delta c_{s1}^2$ by the additional factor $\delta {\cal C}_3$, 
so the deviations of $c_{s2}^2$ and $c_{s3}^2$ from the values 
(\ref{csrad}) and (\ref{cH3}) are very small in the regime (i).

\begin{figure}
\begin{center}
\includegraphics[height=3.3in,width=3.2in]{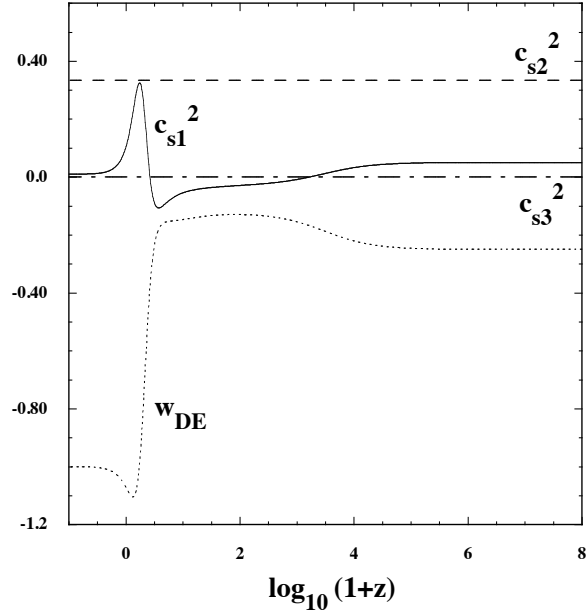}
\end{center}
\caption{\label{fig1}
Evolution of the scalar propagation speed squares 
$c_{s1}^2, c_{s2}^2, c_{s3}^2$ and the dark energy 
equation of state $w_{\rm DE}$ versus the 
redshift $z=1/a-1$ in Model (B).
We choose the model parameters $\alpha=0.3$ and 
$\beta=0.14$ with the initial conditions 
$r_1=5 \times 10^{-11}$, $r_2=8 \times 10^{-12}$, 
$\Omega_r=0.999995$, and $w_m=10^{-3}$ 
at $z=6.0 \times 10^{8}$. 
This case corresponds to the late-time tracking solution 
that approaches the tracker ($r_1=1$) at low redshifts.
During most of the radiation and matter eras
the solution is in the regime $r_1 \ll 1$ and $r_2 \ll 1$, 
in which case the first propagation speed square 
is given by $c_{s1}^2 \simeq (3\Omega_r-1)/40$.}
\end{figure}

Evaluating the term $\xi_2$ for non-relativistic 
matter along the tracker ($r_1=1$), the correction 
$\delta c_{s1}^2$ to $c_{s{\rm H}1}^2$ after the onset of 
the matter-dominated epoch is given by 
\be
\delta c_{s1}^2 \simeq \frac{3(\alpha-2\beta)(1-r_2)}
{(2-3\alpha+6\beta)(1+r_2)}\,.
\ee
In the regime (ii) we have $\delta c_{s1}^2 
\simeq 3(\alpha-2\beta)/(2-3\alpha+6\beta)$, 
while $\delta c_{s1}^2$ vanishes at the de Sitter point (iii). 
{}From Eq.~(\ref{delcH2}) the correction $\delta c_{s3}^2$ to the 
matter sound speed square $c_{s{\rm H}3}^2\,(=O(w_m) \ll 1)$ 
on the tracker can be estimated as
\be
\delta c_{s3}^2 \simeq
\frac{c_{s{\rm H3}}^2}{c_{s{\rm H3}}^2-c_{s{\rm H1}}^2
-\delta c_{s1}^2} 
\frac{9(\alpha-2\beta)^2 r_2 \Omega_m}
{2[2+3r_2(\alpha-2\beta)](1+r_2)(2-3\alpha+6\beta)}\,,
\label{delcs3f}
\ee
which is suppressed both in the regimes (ii) and (iii). 
The correction $\delta c_{s3}^2$ can provide some 
contribution to $c_{s{\rm H3}}^2$ around $r_2=O(0.1)$, 
but $\delta c_{s3}^2$ is still much smaller than 1 
due to the multiplication of the small term 
$c_{s{\rm H}3}^2$ in Eq.~(\ref{delcs3f}).

In Fig.~\ref{fig1} we plot the evolution of the scalar propagation 
speed squares $c_{s1}^2$, $c_{s2}^2$, and $c_{s3}^3$
for the initial conditions $r_1 \ll 1$ and $r_2 \ll 1$ in the 
deep radiation-dominated epoch.
In this case, the dark energy equation of state $w_{\rm DE}$ 
starts to evolve from $-1/4$ (radiation era) to $-1/8$ (matter era) 
and then it reaches a minimum $-1.1$ around the redshift $z=0.3$. 
This late-time tracking behavior is consistent with the observational data 
of Sn Ia, CMB, and BAO at the background 
level \cite{Nesseris}. 

{}From Fig.~\ref{fig1} we find that the first propagation speed 
square $c_{s1}^2$ evolves from the value $1/20$ in the radiation era, 
which is followed by the decrease to the value close to $-1/40$ 
in the matter era. The solution stays in the regime (i) during most 
of the matter-dominated epoch.
The period during which the solution is in the 
regime (ii) is short, so $c_{s1}^2$ soon approaches the value 
$9.7 \times 10^{-3}$ at the de Sitter fixed point (iii) after its 
temporal variation around $z \lesssim O(1)$.
Figure \ref{fig1} shows that the sound speed squares 
$c_{s2}^2$ and $c_{s3}^3$ of radiation and non-relativistic matter 
are close to the values $1/3$ and $0$, respectively. 
This result is consistent with the analytic estimation given above. 
For the parameters used in the numerical simulations of Fig.~\ref{fig1}, 
we find that the deviation of $c_{s2}^2$ from the value $1/3$ is 
less than the order of $10^{-4}$ in the matter era.


\section{Conclusions}
\label{consec} 

We have studied the cosmology of the recently 
proposed generalized Horndeski theories on the flat FLRW background. 
The Lagrangian of these theories is simply expressed in 
terms of three-dimensional scalar quantities constructed 
in the 3+1 ADM decomposition of space-time. 
In Horndeski theories there are particular relations 
(\ref{Hocon}) between the coefficients $A_i$ and $B_i$, 
but GLPV theories are not subject to this restriction.
On the isotropic cosmological background, the perturbation 
equations of motion in GLPV theories are of 
second order with one scalar degree of freedom.

In the presence of multiple scalar fields $\phi_I$ 
($I=1,2,\cdots,N-1$) described by the Lagrangians $P^{(I)}(X_I)$, 
we have expanded the action (\ref{Lagper}) up to 
quadratic order in perturbations of the ADM scalar quantities.
We have in mind the application to dark energy with additional 
perfect fluids of radiation and non-relativistic matter.
The second-order action for tensor perturbations is given by 
Eq.~(\ref{Sh}) with the propagation speed square
$c_t^2={\cal E}/L_{\cal S}$, so the tensor ghosts and 
Laplacian instabilities are absent for $L_{\cal S}>0$ 
and ${\cal E}>0$.
We have derived the second-order Lagrangian density 
for scalar perturbations of the form (\ref{L2mat}), which 
explicitly shows the absence of derivatives higher than 
second order.

The positivity of the $N \times N$ matrix ${\bm K}$ implies 
that the scalar ghosts do not appear under 
the conditions (\ref{scagh1}).
The scalar propagation speeds $c_s$ obey the algebraic 
equation (\ref{cseq2}). 
In Horndeski theories this equation can be written as 
the separate form (\ref{cHeq}), so the solutions 
to $c_s^2$ are simply given by Eqs.~(\ref{cH1})-(\ref{cH2}). 
In GLPV theories the propagation speed squares are 
coupled each other in the form (\ref{ceq}), whose 
right hand side vanishes in the Horndeski limit. 
Under the condition that the deviation of the term ${\cal C}_3$ 
from the Horndeski value $-16L_{\cal S}^2/{\cal W}$ is small, 
we have estimated the propagation speeds in 
Eqs.~(\ref{cs1so})-(\ref{delcH2}).
Compared to the modification $\delta c_{s1}^2$ to the first 
sound speed square $c_{s{\rm H}1}^2$ associated 
with the dark energy field $\chi$, the corrections 
$\delta c_{sI}^2$ ($I=2,3,\cdots, N$) 
to the matter sound speed squares $c_{s{\rm H}I}^2$ 
are generally suppressed.

We have applied our results in Sec.~\ref{persec} to the cosmology 
based on the covariantized Galileon (a class of GLPV theories) and 
the covariant Galileon  (a class of Horndeski theories) in 
the presence of perfect fluids of radiation and 
non-relativistic matter. These two theories give rise to the 
background equations of motion exactly the same as 
each other, so we cannot distinguish them 
at the background level. 

At the level of perturbations, however, different choices of the functions 
$B_4$, $B_5$ give rise to different values of $\cal E$, ${\cal C}_3$ defined 
respectively by Eqs.~(\ref{Edef}) and (\ref{C3def}). As a consequence,
the first scalar propagation speed squares $c_{s{\rm H}1}^2$ in 
Eq.~(\ref{cH1}) differ in these two theories. 
Moreover, in GLPV theories, there is a correction term $\delta c_{s1}^2$ to 
$c_{s{\rm H}1}^2$ estimated approximately by Eq.~(\ref{delcH1}). 
Indeed the first scalar propagation speed square $c_{s1}^2$ 
in the covariantized Galileon becomes negative ($-1/40$) 
in the deep matter era for late-time tracking solutions, while 
in the covariant Galileon $c_{s1}^2=c_{s{\rm H}1}^2=1/40$.
Hence the former is plagued by the small-scale 
instability problem of dark energy perturbations, 
while the latter has a theoretical 
consistent parameter space. 
The matter sound speed 
squares of radiation and non-relativistic matter 
for the covariantized Galileon are close to the values 
(\ref{cH2}) in the Horndeski limit.

We have thus provided a general scheme for studying 
the evolution of background and perturbations in dark energy 
models based on GLPV theories. 
These results will be useful in both placing model-independent 
constraints on the properties of dark energy/modified gravity
and in imposing bounds on individual models. For the latter, 
it may be of interest to search for theoretically and observationally 
allowed parameter spaces in the covariantized version of 
the extended Galileon scenario advocated 
in Refs.~\cite{deFe,extendedGa}.

\section*{Acknowledgements}

This work is supported by the Grant-in-Aid for Scientific 
Research from JSPS (Nos.~24$\cdot$6770 (RK), 24540286 (ST)) 
and by the cooperation programs of Tokyo University of Science and 
CSIC. We thank Antonio De Felice, L\'{a}szl\'{o} \'{A}. Gergely, 
and Shinji Mukohyama for useful discussions. 


\appendix
\section{The autonomous equations in two Galileon theories}
\label{sec:eomhorn}

In both Models (A) and (B) described by the functions 
(\ref{cova1}) and (\ref{cova2}), the variables $r_1$, $r_2$, 
and $\Omega_r$ obey the following equations of motion
\begin{eqnarray}
\label{eq:DRr1}
\hspace{-0.7cm}
r_1' &=& \frac{1}{\Delta} \left(r_1-1\right)
r_1 \left[ r_1 \left(r_1 (-3 \alpha +4 \beta -2)
+6 \alpha -5 \beta \right)-5 \beta \right] \nonumber\\
&&{}\times \left[ 2 \left(\Omega _r+9\right)
+3 r_2 \left( r_1^3 (-3 \alpha +4\beta -2)+
2 r_1^2 (9 \alpha -9 \beta +2)-15 r_1 \alpha 
+14 \beta \right)\right]\,,\\
\label{eq:DRr2}
\hspace{-0.7cm}
r_2' &=& -\frac{1}{\Delta}
[ r_2 (6 r_1^2 (r_2 (45 \alpha ^2-4 (9 \alpha +2) \beta 
+36 \beta^2)-(\Omega_r-7) (9 \alpha -9 \beta +2))
+r_1^3 (-2(\Omega_r+33) (3 \alpha -4 \beta +2) \nonumber \\ & &
-3 r_2 (-2 (201 \alpha +89) \beta +15\alpha
(9 \alpha +2)+356 \beta ^2))
-3 r_1 \alpha (-28 \Omega _r+123 r_2
\beta +36)+10 \beta (-11 \Omega _r
+21 r_2 \beta -3) \nonumber \\ 
& &+3r_1^4 r_2 (9\alpha ^2-30 \alpha (4 \beta +1)+2 (2-9 \beta )^2) 
+3r_1^6 r_2 (3 \alpha -4 \beta+2)^2+3 r_1^5 r_2 
(9 \alpha -9 \beta +2) (3 \alpha -4 \beta +2))], \\
\label{eq:DRr3}
\hspace{-0.7cm}
\Omega_r' &=& \frac{2}{\Delta} \Omega _r [r_1^2 (4(\Omega_r-1) 
(9 \alpha -9 \beta +2)+6 r_2 (-15 \alpha ^2+
36 \alpha \beta +4 (2-9 \beta) \beta ))
-2r_1^3 ((\Omega _r-1) (3 \alpha -4 \beta +2) \nonumber \\
& & +9 r_2 (18 (\alpha +1) \beta
+\alpha  (9 \alpha +2)-36 \beta ^2 ))
+12 r_1 \alpha (-3 \Omega_r+22 r_2
\beta +3)-10 \beta (-4 \Omega_r+21 r_2 \beta +4) 
\nonumber \\
& &+r_1^4 r_2 (549
\alpha ^2+\alpha  (330-840 \beta )+2 (2-9 \beta )^2)
+3 r_1^6 r_2 (3 \alpha -4 \beta
 +2)^2-12r_1^5 r_2 (9 \alpha -9 \beta +2)
(3 \alpha -4 \beta +2)], \nonumber \\
\end{eqnarray}
where 
\begin{eqnarray}
\Delta &\equiv & 2 r_1^4 r_2 [ 72 \alpha ^2+30 \alpha  
(1-5 \beta )+(2-9 \beta )^2 ]+4 r_1^2 
[ 9 r_2 (5 \alpha ^2+9 \alpha \beta +(2-9 \beta ) \beta )
+2 (9 \alpha -9 \beta +2) ] \nonumber\\
&&+4 r_1^3 [ -3 r_2 \left(-2 (15 \alpha +1) \beta 
+3 \alpha  (9 \alpha +2)+4 \beta ^2\right)-3 \alpha 
+4 \beta -2]-24 r_1 \alpha  (16 r_2 \beta +3)
+10\beta (21 r_2 \beta +8).
\end{eqnarray}

Note that we have used the Maple software for deriving these 
autonomous equations and the Fortran software for solving them 
numerically.


\end{document}